\documentclass[aps,prl,twocolumn,showpacs,groupedaddress,longbibliography]{revtex4}

\usepackage[T1]{fontenc}
\usepackage[english]{babel}
\usepackage{epsfig}
\usepackage{graphicx}
\usepackage{color}
\usepackage{amsmath}
\usepackage{float}
\usepackage{bm}
\usepackage{amsbsy}

\begin{document}

\title{Local-field Theory of the BCS-BEC Crossover}

\author{Pietro Maria Bonetti}
\affiliation{Department of Physics "Enrico Fermi", I-56127 Pisa, Italy}
\affiliation{Max Planck Institute for Solid State Research, D-70569 Stuttgart, Germany}

\author{Maria Luisa Chiofalo}
\affiliation{Department of Physics "Enrico Fermi" and INFN, University of Pisa, I-56127 Pisa, Italy}
\affiliation{JILA, University of Colorado, 440 UCB, Boulder, Colorado 80309, USA}
\affiliation{Kavli Institute for Theoretical Physics, University of California, Santa Barbara, CA 93106-4030, USA}

 \email{maria.luisa.chiofalo@unipi.it}

\begin{abstract}
We develop a self-consistent theory unifying the description of a quantum Fermi gas in the presence of a Fano-Feshbach resonance in the whole phase diagram ranging from BCS to BEC type of superfluidity and from narrow to broad resonances, including the fluctuations beyond mean field. Our theory covers a part of the phase diagram which is not easily accessible by Quantum Monte Carlo simulations and is becoming interesting for a new class of experiments in cold atoms.
\end{abstract}

\date{\today}

\pacs{67.85.-d}{Ultracold gases, trapped gases}
\pacs{03.75.Ss}{Degenerate Fermi gases}
\pacs{67.10.-j}{Quantum fluids: general properties}

\maketitle

Quantum gases keep building up considerable interest, as combined experimental-theoretical platforms where the borders between condensed matter, fundamental physics and cosmology can be crossed, with mutual fertilization under the extremely controlled experimental settings and microscopic modeling of atomic physics~\cite{Carr}. Bright examples include a new class of precision measurements~\cite{PezzeSmerziRMP}, Hamiltonian coding inspired by Feynman's idea of quantum simulators~\cite{IBloch,Zohar,VuleticGreinerLukin} for real-time dynamics~\cite{Zohar,ZollerBlatt}, and the quantum phases of Bose/Fermi-Hubbard models relevant to condensed matter~\cite{EDemler,MGreiner,Esslinger}.

Among the intersecting concepts, unexpectedly central remains the paradigm of the crossover from Bose-Einstein Condensation (BEC) to Bardeen-Cooper-Schrieffer (BCS) type superfluidity~\cite{STRINATI_review,Levin_review}. Developed by Leggett~\cite{Leggett} and Nozi{\`e}res and Schmitt-Rink~\cite{Nozieres1985}, its relevance to high-temperature superconductivity (HTSC) was pointed out by Uemura~\emph{et al.}~\cite{Uemura} in a celebrated universal plot, explained in terms of the correlation length~\cite{PistolesiStrinati}, and has become a timely concept for the quantum chromodynamics phase diagram~\cite{GBaym:LCooper} and the equation of state in neutron stars~\cite{PethickNS,Pethick}.

\begin{figure}[htbp]
\centering
\includegraphics[width=0.9\columnwidth]{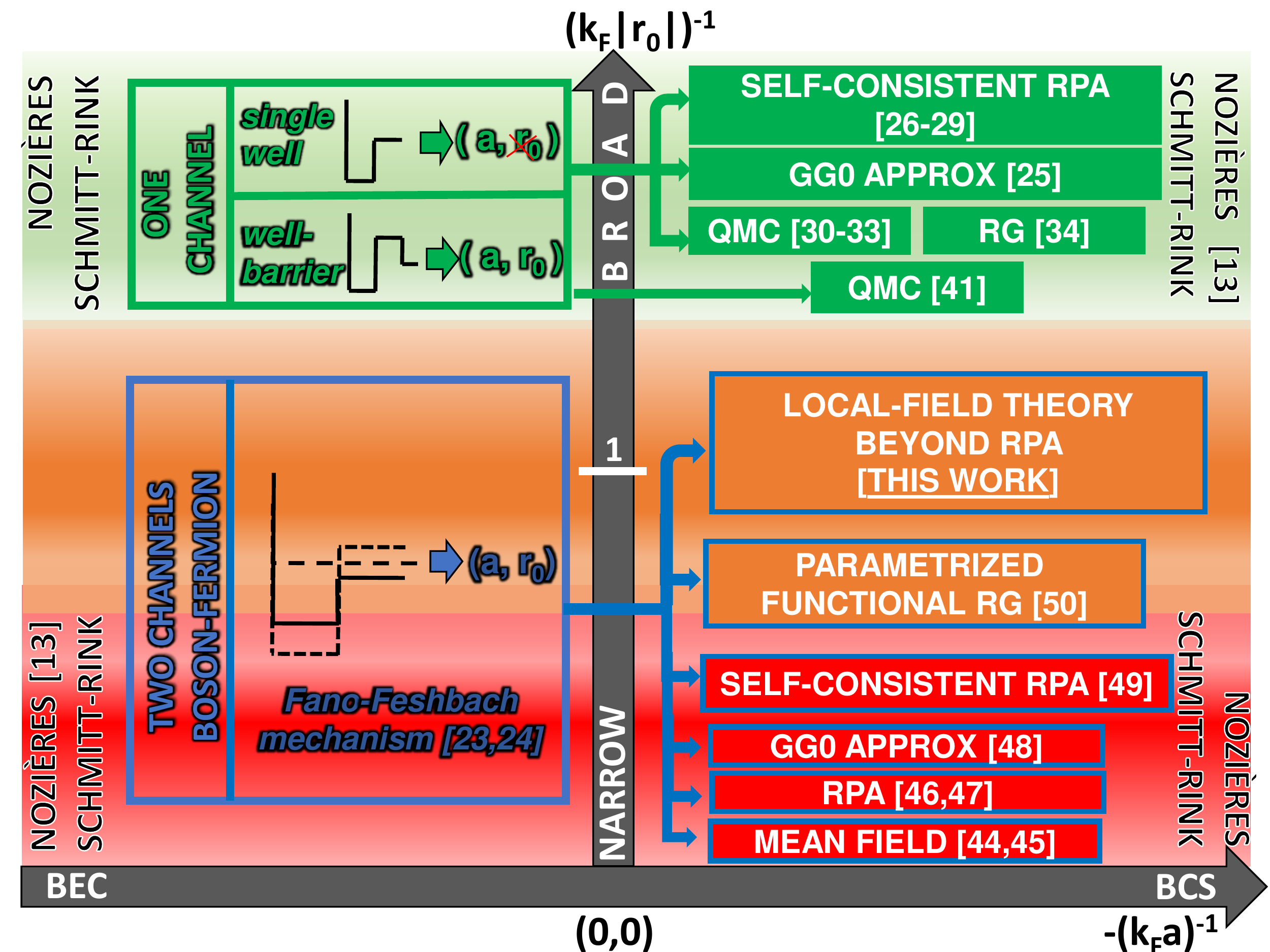}
\caption
    {Conceptual map of BEC-BCS crossover theories in the relevant parameter space defined by $-(k_Fa)^{-1}$, driving the crossover between BEC and BCS limits, and $(k_Fr_0)^{-1}$, driving the resonance width from narrow to broad~\cite{GURARIE}. Sketched are the general model-frameworks (left), i.e. one or two-channel, and the theoretical or Quantum Monte Carlo (QMC) methods (right) used to explore the crossover in narrow (red stripe), intermediate (orange), and broad (green) region (see text). This work bridges the gap of intermediate-to-large values of $(k_Fr_0)^{-1}$, including fluctuations via a unifying local-field theory of the boson-fermion hamiltonian.}
\label{fig:1}
\end{figure}

The advent of Fermi gases~\cite{Jin,Ketterlecrossover,Salomon,Jochim} has turned the crossover physics from a phenomenological approach to gain insight on microscopic theories, into a paradigm to be explored under microscopic mechanisms. Among the latter is the Fano-Feshbach (FF) resonance concept~\cite{FANO,FESHBACH}, where scattering length $a$ and contact interaction strength $U=4\pi\hbar^2a/m$ can be varied at will. The resonance originates from the coupling between a free scattering state of two atoms (open channel) and their bound (closed channel) state (see Fig.~\ref{fig:1}). FF resonances can be classified as narrow (broad) depending on the coupling strength being weak (strong) on the Fermi energy scale $\varepsilon_F\equiv\hbar^2k_F^2/(2m)$. Alternatively, the energy dependence of scattering processes can be embodied in the effective range $r_0$ of the interactions, so that narrow (broad) resonances imply $k_F|r_0|\gg 1$ ($\ll 1$).

\begin{figure}[htbp]
\centering
\includegraphics[width=.7\columnwidth]{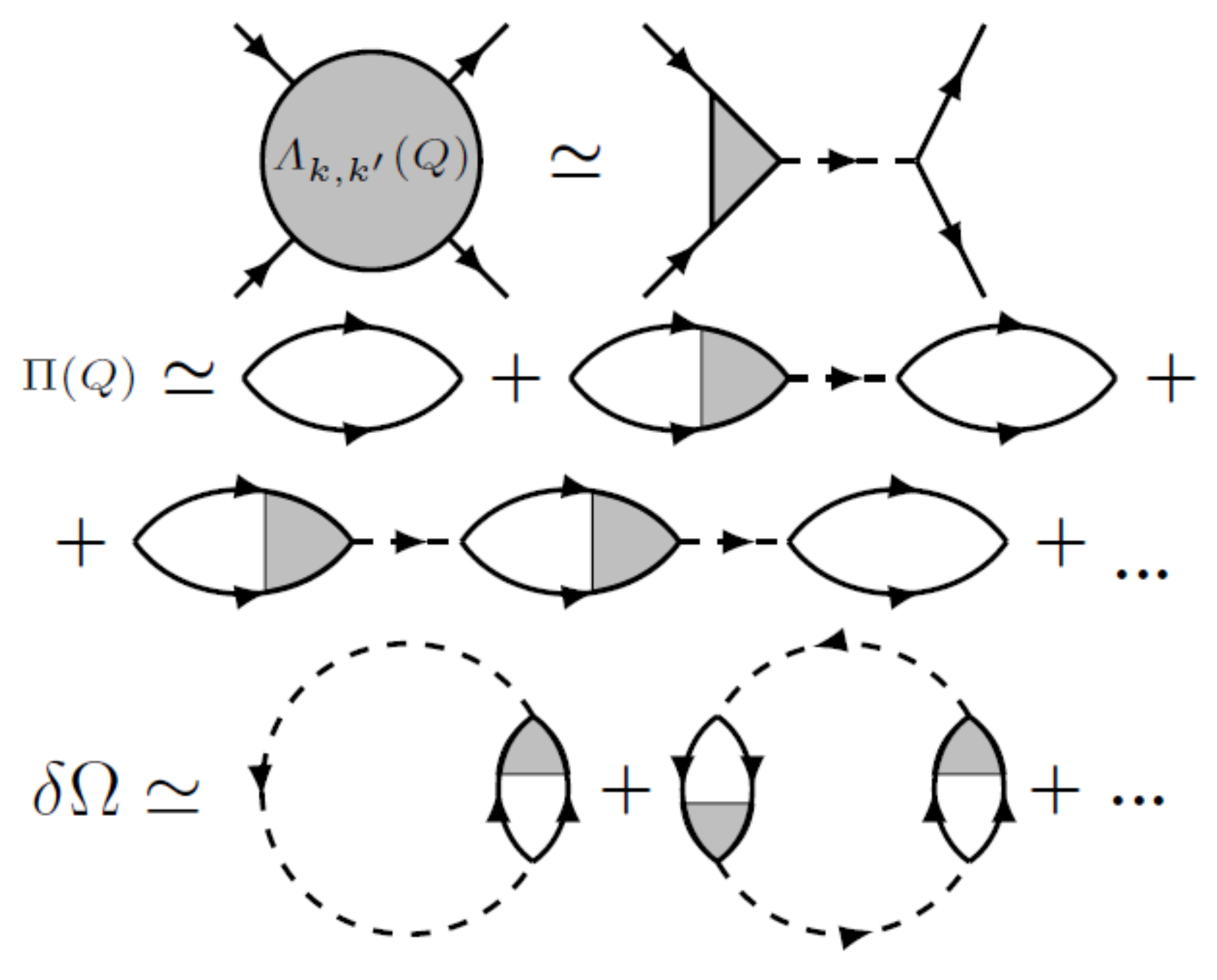}
\caption
    {Local-field approximation: Feynman diagrams picture. Upper row: two-particle irreducible vertex in the particle-particle channel. Dashed line: exchange of a bare boson plus  background interaction. Shaded triangle: renormalized vertex correction, depending on one frequency and one momentum. Central row: pairing susceptibility \eqref{eq: pairing suscept} as an RPA resummation of bubble diagrams connected by two-particle irreducible vertices. Lower row: RPA-like diagrams for the grand-canonical potential.}
\label{fig:2}
\end{figure}

The conceptual map in Fig.~\ref{fig:1} summarizes theories developed so far in this scenario for cold gases. One-channel models build on the BCS Hamiltonian using $U$ as unique parameter, thus suited to describe broad resonances, where the interparticle spacing is the only relevant parameter. Broad resonances have been the norm so far in experiments, very well explored via self-consistent theories including pairing fluctuations~\cite{Levin98,Haussmann1993,PieriStrinati,Strinati_prl,Haussmann_thermo} and Quantum Monte Carlo (QMC) simulations at zero and finite temperature~\cite{Astrakharchik,Bulgac,Troyer,Trivedi} and by Renormalization Group methods \cite{Sachdev}. Intermediate resonances are becoming available in quantum gases experiments~\cite{FFerlaino,greiner,Grimm2011}. Besides, superfluidity in neutron stars~\cite{PethickNS}, is characterized by $k_F|r_0|\simeq 1$. Their theoretical treatment, however, still leaves a number of open questions, stemming from the need of encapsulating the finite width as a second parameter~\cite{GURARIE,JasonHo,ChinGrimm}. Though QMC results are available~\cite{Gandolfi} in a one-channel model mimicking the finite width via well-barrier potentials (Fig.~\ref{fig:1}), they are limited to $(k_F|r_0|)^{-1}\gg 1$. Two-channel, boson-fermion (BF) models instead explicitly include the resonant (boson) state composed by two fermions, embodying the original FF mechanism. Introduced in the HTSC context~\cite{FriedbergLee,Ranninger}, the BF model has been proposed for ultracold atoms in a mean-field formulation~\cite{TIMMERMANS1999,BFmodelHolland}, developed within a Random-Phase-Approximation (RPA)~\cite{Ohashi_Griffin_goldstone,OGprl}, and upgraded to different forms of self-consistent RPA~\cite{Stajic_Levin,LiuHu}. Inclusion of particle-hole fluctuations suited to treat a wide range of FF resonance widths, has been performed within the powerful Functional Renormalization Group (FRG) approach, though in a parametrized manner~\cite{Diehl_ph_fluct,Diehl_renor,Diehl_univers}.   
As a matter of facts, the intermediate regime bridging from narrow to broad FF resonance is devoided of simulational methods and largely unexplored by unifying theoretical methods that include fluctuations beyond mean field.\\
Here we contribute to fill up this theoretical gap. At variance with~\cite{Diehl_ph_fluct}, we develop a theory hinging on a single approximation. With respect to the largely explored broad limit, we predict sizeable effects in $T_c$ at intermediate resonance widths, now accessible in current experiments~\cite{FFerlaino,greiner,Grimm2011}, both at unitarity and in the BCS limit. We take inspiration from local-field dielectric theories~\cite{giuliani2005quantum}, in particular  Singwi-Tosi-Land-Sj\"olander (STLS)~\cite{STLS,HaseShimi} formalism, developed in the 70s to describe the low-density normal electron liquid. In our theory, the superfluid-state symmetries are naturally built in and the Gor'kov and Melik-Barkhudarov screening corrections~\cite{GorkovMelik} recovered. We discuss applications to current experiments and its potential extensions to describe exotic phases away from the superfluid phase.\\
\noindent \textit{The theory-}
We consider the boson-fermion (BF) grand-canonical Hamiltonian \cite{FriedbergLee}: 
\begin{equation}
\begin{split}
\mathcal{H}-\mu \,\mathcal{N}&=\sum_{\mathbf{k},\sigma}\varepsilon_{\mathbf{k}}\,c^{\dagger}_{\mathbf{k},\sigma}c_{\mathbf{k},\sigma}+\sum_{\mathbf{q}}\varepsilon_{\mathbf{q}}^B\,b_{\mathbf{q}}^{\dagger}b_{\mathbf{q}}+\\
+&\frac{g}{\sqrt{V}}\sum_{\mathbf{k},\mathbf{q}}\left[b^{\dagger}_{\mathbf{q}}\,c_{\mathbf{q}/2-\mathbf{k},\downarrow}\,c_{\mathbf{q}/2+\mathbf{k},\uparrow}+\text{h.c.}\right]+\\
-&\frac{U_{bg}}{V}\sum_{\mathbf{k},\mathbf{k}',\mathbf{q}}c^\dagger_{\mathbf{q}/2+\mathbf{k},\uparrow}\,c^\dagger_{\mathbf{q}/2-\mathbf{k},\downarrow}\,c_{\mathbf{q}/2-\mathbf{k}',\downarrow}\,c_{\mathbf{q}/2+\mathbf{k}',\uparrow}.
\end{split}
\label{eq: BF hamiltonian}
\end{equation}
The operator $c^{\dagger}_{\mathbf{k},\sigma}$ ($b_{\mathbf{k}}^{\dagger}$) creates a spin-1/2 fermion (spinless boson) with momentum $\mathbf{k}$. 
The first two terms represent the fermion and boson kinetic energies $\varepsilon_{\mathbf{k}}=k^2/(2m)-\mu$ and $\varepsilon_{\mathbf{q}}^B=q^2/(4m)-2\mu+2\nu$, in terms of the fermionic mass $m$, chemical potential $\mu$ and energy $2\nu$ of the resonant state. The factors $4m$ and $2\mu$ in the dispersion account for the bosons being composed of two fermions. $2\nu$ is the crucial parameter driving the system from the Fermi limit at large detunings $\nu\gg\varepsilon_F$, where bosons exist only as virtual states, to the pure Bose limit $\nu$$\ll$$-\varepsilon_F$ with a real macroscopic occupation of the resonant state. Bosons and fermions are hybridized via the coupling with strength $g$, converting two fermions into a boson and viceversa, related to the effective range $r_0$ of the scattering potential via $r_0=-8\pi\hbar^4/(mg)^2$ \cite{GURARIE,Kokkelmans_Renorm}. The theory embodies two independent physical parameters, $k_Fr_0$ and $k_Fa$, in the model tuned via $g$ and $\nu$. To which the background scattering length $a_{bg}$ ($U_{bg}\equiv 4\pi\hbar^2 a_{bg}/m$) joins to account for scattering away from resonance.\\
\textit{The method-}
Our method hinges on the concept of local field in dielectic-function theories, introduced to study the density and spin response of the electron liquid in low-density metals, where the Coulomb interaction dominates over the kinetic energy~\cite{Hubbard,STLS}. While referring to the Supplemental Material (SM)~\cite{SoM} for details, here is the concept essence. The system response is determined by introducing the exchange and correlation (xc) potential $V_{\text{xc}}(\vec{r},t)=-\int dt'd\vec{r}\,v(|\vec{r}-\vec{r}'|)G_{L}(|\vec{r}-\vec{r}'|,t-t')\delta n(\vec{r}',t')$, in terms of the so-called local-field factor $G_{L}$ generated by the polarization density locally induced in the medium and describing the hole dug in around a given particle by xc processes. In the STLS scheme~\cite{STLS}, $G_L$ is determined by the xc-generalized force driven by the fluid and weighted over the static probability of finding a particle at distance $r$,  measured by the pair-correlation function $g(\vec{r})$ \cite{giuliani2005quantum,SoM}. The equations set is closed by relating $g(\vec{r})$ to the structure factor, and the latter to the imaginary part of the response via the fluctuation-dissipation theorem. In different language, the choice of $G_{L}$ amounts to define the irreducible interaction determining the vertex corrections. Inspired by these physical ideas, we now turn to implement them in the BF theory. \\
As detailed in the SM ~\cite{SoM}, our method naturally embodies the spin-$SU(2)$ and time-reversal symmetries dictated by the Hamiltonian ~\eqref{eq: BF hamiltonian}, and those emerging from gauge transformations, like the Hugenoltz-Pines theorem~\cite{Hugenh-Pines} ensuring that the excitation spectrum be gapless while the Goldstone mode sets in. The resulting complex formalism can be represented in a quite compact form, but in order to comprehensively reveal the essence of the theory, we reduce it to a minimum by first focusing on the calculation of the superfluid transition temperature $T_c$.\\
Since the Thouless criterion states that the divergence of the the pairing susceptibility $\Pi\left(\mathbf{q},\omega\right)$ is related to the divergence of the particle-particle scattering vertex, we start by evaluating  $\Pi\left(\mathbf{q},\omega\right)=i\int_{0}^{\infty}dt\,e^{i\omega t}\, \langle 
{\mathcal{C}^\dagger_{\mathbf{q}}\left(t\right)}{\mathcal{C}_{\mathbf{q}}\left(0\right)} \rangle$,
where the operator $\mathcal{C}_{\mathbf{q}}\equiv\sum_{\mathbf{k}}\left[c_{-\mathbf{k}+\mathbf{q}/2,\downarrow}c_{\mathbf{k}+\mathbf{q}/2,\uparrow}\right]$ annihilates a fermion pair with total momentum $\mathbf{q}$ and averages are meant at equilibrium as in linear response. 
We perturb the pairing fields by acting with the source term $J_{\mathbf{q}}(t)$ explicitly breaking the $U(1)$ symmetry, i.e. adding $\sum_{\mathbf{q}}\left[J^*_{\mathbf{q}}(t)\mathcal{C}_{\mathbf{q}}(t)+J_{\mathbf{q}}(t)\mathcal{C}^\dagger_{\mathbf{q}}(t)\right]$ to \eqref{eq: BF hamiltonian}. We then compute the system linear response by the equation of motion method. In fact, the generalized Wigner distribution function 
$f^*_{\mathbf{k},\mathbf{q}}(t)=\langle c^\dagger_{\mathbf{k}+\mathbf{q}/2,\uparrow}(t)c^\dagger_{-\mathbf{k}+\mathbf{q}/2,\downarrow}(t)\rangle$,
turns out to be more practical to work with, than $\mathcal{C}^{(\dagger)}$. After Fourier transforming in frequency domain, we obtain:
\begin{eqnarray}
&&\omega f^*_{\mathbf{k},\mathbf{q}}=\left(\varepsilon_{\mathbf{k}+\mathbf{q}/2}+\varepsilon_{\mathbf{k}-\mathbf{q}/2}\right)f^*_{\mathbf{k},\mathbf{q}}
\nonumber \\
&&+\sum_{\mathbf{q}',\sigma}
\left(\frac{1}{2}\delta_{\mathbf{q},\mathbf{q}'}-\big\langle\mathcal{D}^\sigma_{\text{sgn}\sigma\mathbf{k}+\mathbf{q'}/2,\mathbf{q}-\mathbf{q}'}\big\rangle\right)J^*_{\mathbf{q}'}(\omega)\\
&&+\sum_{\mathbf{q}',\sigma}\bigg\langle\left(\frac{1}{2}\delta_{\mathbf{q},\mathbf{q}'}- \mathcal{D}^\sigma_{\text{sgn}\sigma\mathbf{k}+\mathbf{q'}/2,\mathbf{q}-\mathbf{q}'}\right) U_{\text{eff}}(\mathbf{q}',\omega)\mathcal{C}^{\dagger}_{\mathbf{q}'}\bigg\rangle,\nonumber  
\label{eq: EOM wigner}
\end{eqnarray}
where $\mathcal{D}^\sigma_{\mathbf{k},\mathbf{q}}\equiv c^\dagger_{\mathbf{k}+\mathbf{q}/2,\sigma}c_{\mathbf{k}-\mathbf{q}/2,\sigma}$, $\text{sgn}(\uparrow)=+1$, $\text{sgn}(\downarrow)=-1$, and $\delta_{\mathbf{q},\mathbf{q}'}\equiv(2\pi)^3\delta^3(\mathbf{q}-\mathbf{q}')$. The effective interaction 
$U_{\text{eff}}(\mathbf{q},\omega)=-U_{bg}+g^2D_0(\mathbf{q},\omega)=-U_{bg}+{g^2}/({\omega-\varepsilon^B_{\mathbf{q}}})$ is driven by the bare contact and the exchange of a resonant boson. In the above equation, we have omitted the $\omega$-dependence of $f_{\mathbf{k},\mathbf{q}}$ for the sake of simplicity.
Average over the unperturbed system yields $\langle\mathcal{D}^{\alpha}_{\mathbf{k},\mathbf{q}}\rangle=n_{\mathbf{k}}\,\delta_{\mathbf{q},0}$, with $n_{\mathbf{k}}=2T\sum_{i\omega_n}G(\mathbf{k},i\omega_n)e^{i\omega_n0^+}$ the momentum distribution, that can be computed from the fermionic Green's function $G(\mathbf{k},i\omega_n)$ once the self-energy is known. Postponing this task, we begin by approximating  $n_{\mathbf{k}}=f(\varepsilon_{\mathbf{k}})=[{e^{\beta\varepsilon_{\mathbf{k}}}+1}]^{-1}$ with the non-interacting $G_0(\mathbf{k},i\omega_n)=(i\omega_n-\varepsilon_{\mathbf{k}})^{-1}$. The third term is more complicated, being an average of four operators. The equation of motion for it would contain higher-order terms in an infinite hierarchy~\cite{Niklasson}. We close the equation set by generalizing the STLS idea~\cite{STLS} to the pairing channel. To gain physical insight, we revert back to real space and approximate the connected average as $\sum_{\alpha=\pm}\langle\delta^3(\mathbf{r})/2-c_{\sigma(\alpha)}^{\dagger}(\mathbf{x}_{\alpha}) c_{\sigma(\alpha)}(\mathbf{x}_{-\alpha})\rangle g_{\text{corr}}\left({r}/{2}\right)\langle \varrho^+(\mathbf{x}_{-\alpha})\rangle$,
with $\mathbf{x}_{\alpha}\equiv  \mathbf{R}+{\alpha\mathbf{r}}/{2}$.
The core of our approximation is the Cooper-pair correlation function $g_{\text{corr}}(|\mathbf{R}-\mathbf{x}|)\equiv \big\langle \mathcal{C}(\mathbf{R})\,\mathcal{C}^\dagger(\mathbf{x})\big\rangle_0$, describing the correlations occurring whenever a Cooper pair is destroyed at $\mathbf{R}$ and a second one created at $\mathbf{x}$. As an equilibrium average, $g_{\text{corr}}$ depends only on $|\mathbf{R}-\mathbf{x}|$ and not on time.
We remark that this is the only approximation in our theory. At variance with other approaches~\cite{Diehl_ph_fluct}, once  performed all the rest fully consistently follows.\\
Applying the extended STLS decoupling and transforming back to $\mathbf{q}$ space~\cite{SoM}, the static pairing structure factor $S(\mathbf{q})$ naturally appears, related to $g_{\text{corr}}(r)$ by Fourier transform. Solving for $f^*_{\mathbf{k},\mathbf{q}}$ and using $\langle\mathcal{C}^\dagger_{\mathbf{q}}\rangle=\sum_{\mathbf{k}}f^*_{\mathbf{k},\mathbf{q}}=\Pi(\mathbf{q},\omega)J^*_{\mathbf{q}}$, the pairing susceptibility reads~\cite{SoM}:
\begin{equation}
    \Pi(\mathbf{q},\omega)=\frac{\Pi_0(\mathbf{q},\omega)}{1-\left[1-\mathcal{G}(\mathbf{q},\omega)\right]U_{\text{eff}}(\mathbf{q},\omega)\Pi_0(\mathbf{q},\omega)},
    \label{eq: pairing suscept}
\end{equation}
in terms of the non-interacting $\Pi_0(\mathbf{q},\omega)=\sum_{\mathbf{k}}$ $ [{1-f(\varepsilon_{\mathbf{k}+\mathbf{q}/2})-f(\varepsilon_{\mathbf{k}-\mathbf{q}/2})}]/[{\omega-\varepsilon_{\mathbf{k}+\mathbf{q}/2}-\varepsilon_{\mathbf{k}-\mathbf{q}/2}}]$ and
\begin{equation}
    \mathcal{G}(\mathbf{q},\omega)=-\sum_{\mathbf{q}'}\frac{\Pi_0(\mathbf{q},\mathbf{q}';\omega)}{\Pi_0(\mathbf{q},\omega)}S(\mathbf{q}-\mathbf{q}'), 
    \label{eq: loc field fact}
\end{equation}
the \textit{local field factor}. In \eqref{eq: loc field fact}, the function $\Pi_0(\mathbf{q},\mathbf{q}';\omega)$ is obtained after replacing $\mathbf{q}\rightarrow \mathbf{q}'$ only in the numerator of the definition of $\Pi_0(\mathbf{q},\omega)$. Finally, $S(\mathbf{q})$ is related to $\Pi(\mathbf{q},\omega)$ via the fluctuation-dissipation theorem:
\begin{equation}
    S(\mathbf{q})=\int_{-\infty}^{+\infty}\frac{d\omega}{2\pi}\,\frac{2}{1-e^{-\beta\omega}}\Im\left\{\Pi(\mathbf{q},\omega+i\delta)\right\}. 
\label{eq:SB}
\end{equation}
The presence of the static $S(\mathbf{q})$ in $\mathcal{G}(\mathbf{q},\omega)$ can be alternatively derived by extending to the particle-particle channel the Niklasson calculation \cite{Niklasson}: in~\eqref{eq: EOM wigner}, one can write the equations of motion for the last average and show that $S(\mathbf{q})$ appears while $\omega \to +\infty$. Eqs. \eqref{eq: pairing suscept}-\eqref{eq:SB} form a closed set, extending the STLS approach to the presence of a pairing field driven by the microscopic Fano-Feshbach mechanism. Given $T$ and $\mu$, their self-consistent solution provides the pairing susceptibility beyond mean field, and therefore all the fluid properties. 
We now proceed to determine the evolution of $T_c$ in the crossover. The Thouless criterion amounts to require that the denominator in \eqref{eq: pairing suscept} vanishes: $1-\left[1-\mathcal{G}(0,0)\right]U_{\text{eff}}(0,0)\Pi_0(0,0)=0$. i.e. 
\begin{equation}
\left[1-\mathcal{G}(0,0)\right]\left[U_{bg}+\frac{g^2}{2\nu-2\mu}\right]\sum_{\mathbf{k}}\frac{\tanh\left({\beta\varepsilon_{\mathbf{k}}}/{2}\right)}{2\varepsilon_{\mathbf{k}}}=1.
    \label{eq: Tc}
\end{equation}
Notice that~\eqref{eq: Tc} can be viewed as the conventional RPA equation for $T_c$, with the interaction corrected by $\left[1-\mathcal{G}(0,0)\right]$. We will comment on the physics later on. We now need an equation for $\mu$, deriving the corresponding number equation from a diagrammatic argument. Indeed, \eqref{eq: pairing suscept} can be viewed as a RPA resummation of diagrams as in Fig.~\ref{fig:2} (a)-(b), consisting on $N$ bubbles connected by $N-1$ interaction lines,  the latter corresponding to the free boson propagator $D_0(\mathbf{q},\omega)$ plus the bare $U_{bg}$ 
corrected by $\left[1-\mathcal{G}\right]$. In essence, the local-field approximation amounts to estimate the particle-particle irreducible vertex in the pairing channel as $U_{\text{eff}}[1-\mathcal{G}]$. Summing up all the closed ring diagrams 
as in Fig.~\ref{fig:2}(c), we get the interaction correction $\delta\Omega$ to the noninteracting grand-canonical potential $\Omega_0$~\cite{SoM}.
We obtain the same $\delta\Omega$ by the running-coupling constant method, after neglecting the intrinsic dependence of $\mathcal{G}$ on $g$ and $U_{bg}$. Thus, we expect this approximation to be quantitatively reliable for small to intermediate values of $g$ and $U_{bg}$. We then derive $\mu$ from
\begin{equation}
    n=-\frac{\partial}{\partial\mu}(\Omega_0+\delta\Omega).
    \label{eq: mu}
\end{equation}
Eqs.~\eqref{eq: loc field fact}-\eqref{eq: mu} are the closed set describing the critical behavior in the crossover for narrow-to intermediate FF resonances. For their self-consistent solution, one iterates an initial guess for $\mathcal{G}$ (e.g. $\mathcal{G}=0$) until convergence.\\
Once discussed the essence of the theory, we now relax the $G_0$ approximation: we relate the scattering vertex  $\Gamma=\left[\left(1-\mathcal{G}\right)U_{\text{eff}}\right]/\left[1-\left(1-\mathcal{G}\right)U_{\text{eff}}\Pi_0\right]$ to $\mathcal{G}$, the fermionic self-energy 
$\Sigma(K_n)=\sum_{Q_n}\Gamma(Q_n)G_0(Q_n-K_n)$ to $\Gamma(Q_n)$,
then updating $G_0\to G=[G_0^{-1}-\Sigma]^{-1}$ and $\Pi_0\to\widetilde{\Pi}_0(Q_n)=\sum_{K_n}G(Q_n-K_n)G(K_n)$ in the vertex expression, with $n=\sum_{K_n}G(K_n)e^{i\omega_n0^+}$ and the $\mathcal{G}$ constant during the loop. Here, $Q_n=(\nu_n,\mathbf{q})$ ($K_n=(\omega_n,\mathbf{k})$) is the 4-vector with a bosonic (fermionic) Matsubara frequency. This the analogue of a GW approximation~\cite{giuliani2005quantum}.\\  
\noindent \textit{Limiting cases-}
Despite the equations complexity, we can extract relevant analytical limits. We first need to regularize the (otherwise diverging) non-interacting susceptibilities. This requires~\cite{Kokkelmans_Renorm,OhashiGriffinTrap} to renormalize $U_{bg}$, $g$ and $\nu$ into $U_R$, $g_R$ and $\nu_R$, exactly as in the two-body problem~\cite{SoM}. From now on we drop, for simplicity, $U_{bg}$.
In the BCS limit with $\nu_R\gg\varepsilon_F$ and $T_c\ll T_F$
~\eqref{eq: mu} reads $\mu\simeq \varepsilon_F$, so that $T_c^{\text{BCS}}\simeq {8e^{\gamma-2}{\pi}^{-1}\,\varepsilon_F}\exp\left[-({2\nu_R-2\varepsilon_F})/({N(0)\Tilde{g}^2})\right]$, 
where $N(0)$ is the density of states at $\varepsilon_F$ and $\Tilde{g}^2\equiv g_R^2[1-\mathcal{G}(0,0)]$. It can be shown that $0<\mathcal{G}(0,0)<1$ if an RPA $S(\mathbf{q})$ is inserted in~\eqref{eq: loc field fact}. Thus, at fixed $\nu_R$, the local field correction suppresses $T_c$ with respect to its mean-field value. This result is reminiscent of the celebrated Gor'kov and Melik-Barkhudarov (GMB) correction in one-channel calculations~\cite{GorkovMelik}, stating that in the BCS limit particle-hole processes suppress $T_c$ and superfluid gap by a factor $\simeq 2.2$~\cite{StrinatiGMB}. 
In our theory, these particle-hole corrections show up in the $g$ renormalization. Indeed, evaluating in the BCS limit $\nu_R\gg\varepsilon_F$~\cite{STRINATI_review} the BF vertex diagram $\Lambda^{\text{BF}}_{\mathbf{k},\omega}(\mathbf{q},\Omega)$ to lowest order as in Fig.~\ref{fig:3}, we get:
\begin{figure}[ht]
\centering
\includegraphics[width=0.5\columnwidth]{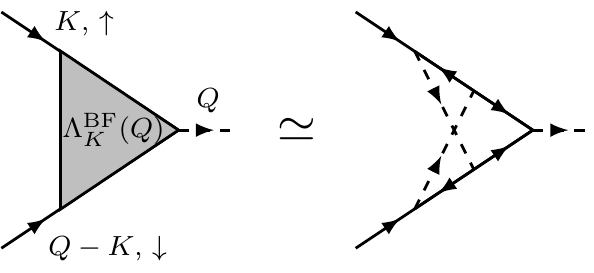}
    \caption{\footnotesize{Lowest order contribution to the irreducible boson-fermion vertex $\Lambda^{\text{BF}}_{K}(Q)$, showing the coupling constant $g$ is renormalized by particle-hole processes (dashed lines are bare bosons). 
    }}
    \label{fig:3}
\end{figure}
\begin{equation}
    \Tilde{g}_{\text{GMB}}^2=g_R^2\left[1-\frac{g_R^2}{2\nu_R-2\varepsilon_F}N(0)\ln(4e)^{1/3}\right].
\end{equation}
Replacing $g_R^2[1-\mathcal{G}]$ by $\Tilde{g}_{\text{GMB}}^2$ in~\eqref{eq: Tc}, $T_c$ results suppressed by $(4e)^{1/3}\simeq 2.2$. From the structure of our equations, using local field factors amounts to neglect the ($\mathbf{k}$, $\omega$) dependence in the full boson-fermion vertex $\Lambda^{\text{BF}}_{\mathbf{k},\omega}(\mathbf{q},\Omega)$, like in electron liquids~\cite{giuliani2005quantum}. This defines the perimeter of our theory in capturing the $T_c$ suppression effects.\\
On the BEC limit with $\nu\ll-\varepsilon_F$, eq.~\eqref{eq: Tc} yields $\mu\simeq\nu$ and from \eqref{eq: mu}
one obtains the BEC $T_c$ of $n/2$ bosons with mass $2m$: $T_{c}^{\text{BEC}}=({\pi}/{m})[{n}/({2\zeta(3/2)})]^{2/3}$.\\
\noindent\textit{Implications for current experiments-} Self-consistent calculations in the narrow-resonance case with $k_F|r_0|\simeq 5$ in the BCS limit, yields a more limited suppression with our theory, so that $T_c$ is enhanced by a factor up to $\simeq 10\%$ than the (perturbative) $T_{c,\text{GMB}}$. The resonance turns out to be characterized by a maximum at $T_c$~\cite{Strinati_maximumTc}, that reduces towards the narrow limit~\cite{tobepub}: in the broad resonance limit with $k_Fr_0\simeq 0.5$ we get $T_c/T_F\simeq 0.22$, comparable with the QMC value $T_c/T_F=0.24 (2)$ by Bulgac et al.~\cite{Bulgac}. At unitarity, varying the resonance width in the range $0.5<k_F|r_0|<5$ by one order of magnitude yields variations of the maximum $T_c$ up to $\simeq 8\%$~\cite{tobepub}.
\\
\noindent \textit{Conclusions-}
We have developed a unifying fully self-consistent theory of superfluidity with pairing fluctuations beyond mean field, hinging on the original Fano-Feshbach microscopic resonant mechanism. Our theory bridges the description of the BCS-BEC crossover from narrow to broad FF resonances, in a region so far devoided of simulational methods and largely unexplored by theoretical methods. We brush up the old-fashioned concept of local field, successfully developed in electron liquids, and demonstrate its so-far unexplored methodological power to access a complex phase diagram where density, spin, and amplitude/phase fluctuations of a superfluid order parameter can be treated on equal footing with only one physical approximation.\\  
Intermediate resonance widths are becoming accessible by a new class of experiments, like with fermionic Er atoms~\cite{Ferlaino}, Fermi-Hubbard simulators~\cite{greiner}, or Fermi-Bose mixtures~\cite{Grimm2011}, that can provide a test-bed for our theory. A systematic study of relevant observables like $T_c$ and superfluid gap at $T\ll T_c$, requires a full numerical solution, that is under way~\cite{tobepub}. Effects up to $10\%$ found in $T_c$ with respect to the broad resonance case, open up unexplored physics. Interest is also building up on the equation of state in neutron stars, where observational data are compatible with $k_Fa\simeq -13$ and $k_F|r_0|\simeq 2$, though on the fully different fm length scale~\cite{GBaym:LCooper}.
\begin{acknowledgments}
We thank Michele Barsanti, who is developing the numerical environment for the theory~\cite{tobepub}. M.L.C. would like to thank JILA and KITP for kind hospitality, while part of this work has been carried out. We would like to thank Murray J. Holland, Eugene Demler, Andrea Perali, Pierbiagio Pieri and Giancarlo Strinati for inspiring discussions. We are also grateful to Walter Metzner for a careful reading of the manuscript. This work is dedicated to Debbie Jin.
\end{acknowledgments}

\bibliographystyle{apsrev}

\end{document}


\title{Supplemental Material for: \\ \mytitle}

\author{Pietro Maria Bonetti}
\affiliation{Department of Physics "Enrico Fermi" and INFN, University of Pisa, I-56127 Pisa, Italy}
\affiliation{Max Planck Institute for Solid State Research, D-70569 Stuttgart, Germany}

\author{Maria Luisa Chiofalo}
\affiliation{Department of Physics "Enrico Fermi" and INFN, University of Pisa, I-56127 Pisa, Italy}
\affiliation{JILA, University of Colorado, 440 UCB, Boulder, Colorado 80309, USA}
\affiliation{Kavli Institute for Theoretical Physics, University of California, Santa Barbara, CA 93106-4030, USA}

\begin{abstract}
This document contains details on local-field factor theories, and the extension of the theory developed in the main text, to the case with finite background scattering length.
\end{abstract}

\maketitle
\section{Local-field factor theories in a nut-shell}
\begin{figure}[htbp]
\centering\includegraphics[width=0.7\textwidth]{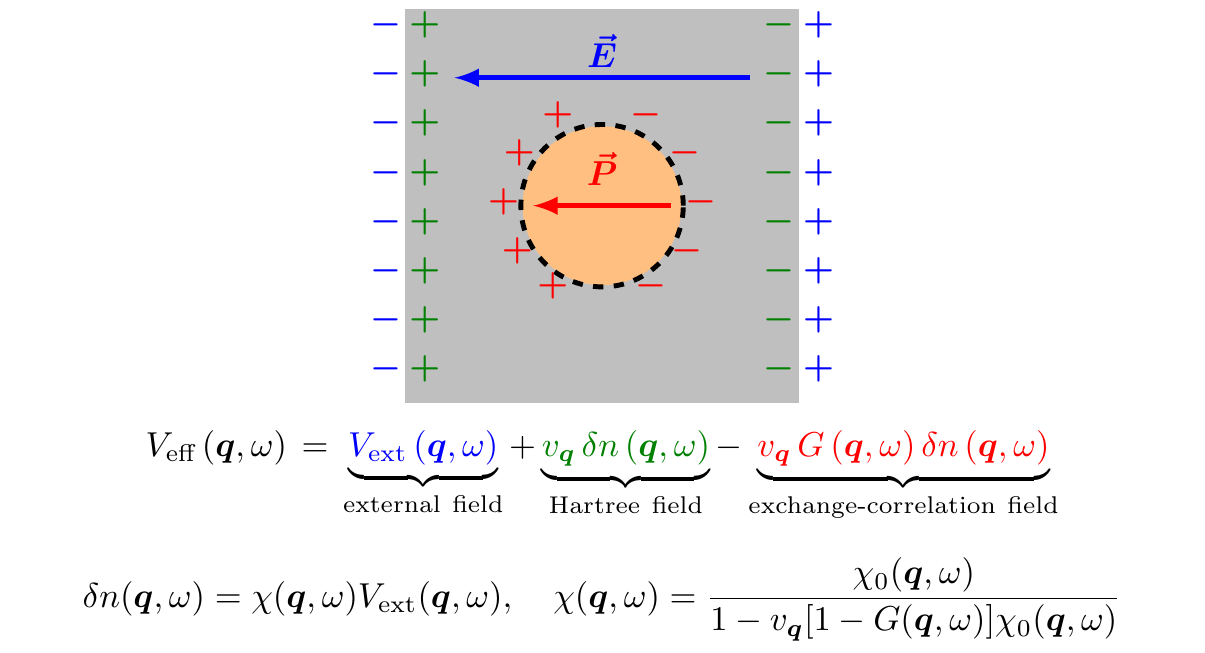}
\caption{Schematic picture of the local-field factor approximation. In a classical perspective, once an electric field is applied to a dielectric material, the field locally felt by a single electron is reduced by the polarization induced by the other electrons surrounding it. This is a pure correlation effect. In a quantum many-body system, the potential felt by an electron is composed by adding up the external, the Hartree, and a local correlation fields. The Hartree field, as a first approximation, describes correlations accounting the presence of the other electrons under a self-consistent mean-field. The correlation field accounts for the fact that the electron itself is actively participating to the creation of the mean-field. In local-field approximation, this last last term is written as a purely local one (in frequency and momentum space).}
\label{fig:1_SM}
\end{figure}

The local-field factor theories for the dielectric function~\cite{giuliani2005quantum} have originally been developed in the 70s to describe the low-density normal electron liquid. We here illustrate the basics of the theory for the simpler case of the density response in a normal liquid, taken to be the electron liquid as in the original formulation. With this strategy, we wish to make easier for the reader tha task of following the formalism for the more complex case treated in the main text and detailed in the next section, that is characterized by many variables (density, spin, and amplitude and phase of the order parameter) and by broken symmetry within a boson-fermion Hamiltonian. 

The essence of local-field factor theories is well understood from the analogy with the electrodynamic response of a medium in a Lorentz cavity, which we now remind. The concept is sketched in Fig.~\ref{fig:1_SM}, which in the following we refer to. While the response of the (non-interacting) system to the external potential $V_{ext}(\vec{r},t)$ driven by a corresponding electric field amounts to the Lindhard approximation, the Random-Phase Approximation (RPA) represents the response to $V_{ext}(\vec{r},t)+V_{H}(\vec{r},t)$, where $V_{H}(\vec{r},t)=\int d\vec{r}v(|\vec{r}-\vec{r}'|)\delta n(\vec{r}',t)$ is the Hartree mean-field potential sourced by the charge $\delta n(\vec{r}',t)$ induced at the cavity boundaries in the presence of the two-body interaction $v(|\vec{r}-\vec{r}'|)$. In fact, RPA is known to badly overestimate the effects of interactions among the particles in the medium, since they are allowed to be closer than they really do. For a beyond-mean field treatment, one therefore has to consider also the potential sourced by the polarization charges that are locally induced in the medium. This can be expressed as $V_{xc}(\vec{r},t)=-\int dt'd\vec{r}v(|\vec{r}-\vec{r}'|)G_{L}(|\vec{r}-\vec{r}'|,t-t')\delta n(\vec{r}',t')$, in terms of the so-called local-field factor $G_{L}$. In a quantum fluid, this in fact describes the effects of the hole surrounding a given particle dug in by exchange and correlation processes. The local field $G_L$ can be related to the structure of the fluid, and the latter - via the fluctuation-dissipation theorem - to the response function expressed in terms of $G_L$. This procedure provides a closed set of equations for the response function of the system, to be solved self-consistently. 

The whole point then amounts to how $G_L$ can be related to the structure of the fluid. A number of approximations have been developed and tested for the electron liquid, starting from the Hubbard approximation, where a form of $G_L$ is built up that interpolates between known infrared and ultraviolet behaviors~\cite{giuliani2005quantum}. A popular approximation has been developed by Singwi, Tosi, Land, and Sj\"olander (STLS)~\cite{STLS} and extended to the quantum regime by Hasegawa and Shimizu~\cite{HaseShimi}. Improvements on the long-wavelength behavior of the STLS theory have been developed by Vashishta and Singwi~\cite{VS72}, to build in the compressibility sum rule by including the density dependence of the pair correlation function. Notwithstanding their simplicity and differences, these models were capable of successfully describing the ground-state properties in remarkable agreement with Quantum Monte Carlo simulations, and have been extended to treat multicomponent and spin-polarized systems, and current and transverse spin response~\cite{giuliani2005quantum} in normal systems. We have taken inspiration from the STLS scheme, that is known to represent an optical trade-off between performance and simplicity for most properties. 

In the STLS scheme, $G_L$ is determined by noting that the generalized force due to exchange and correlation in the fluid can be expressed as 
\begin{equation}
   \vec{\nabla}_{\vec{r}}V_{xc}(\vec{r},t)=\int [g(\vec{r}-\vec{r}')-1]\vec{\nabla}_{\vec{r}}v(|\vec{r}-\vec{r}'|)\delta n(\vec{r}',t'),
   \label{eq:sm_genforce}
\end{equation}
that is after weighting the bare force $\vec{\nabla}_{\vec{r}}v(|\vec{r}-\vec{r}'|)\delta n(\vec{r}',t')$ over the probability $[g(\vec{r}-\vec{r}')-1]$ dictated by the pair-correlation function $g(\vec{r})$ \cite{giuliani2005quantum} and then summing up over all the fluid slices. Notice that the explicit choice has been performed of confining all the dynamical effects into the induced density, so that the static $g(\vec{r})$ is involved. In momentum space, eq. \eqref{eq:sm_genforce} can be recast in terms of the static structure factor, related to the pair correlation function $g(r)$ by Fourier transform:
\begin{equation}
   G(\vec{q})=-\frac{1}{(2\pi)^3n}\int \frac{\vec{q}\cdot\vec{q}'}{q^2}\frac{v(\vec{q}')}{v(\vec{q})}[S(\vec{q}-\vec{q}')-1]d\vec{q}',
   \label{eq:sm_GqSq}
\end{equation}
with $n$ the system density and $v(\vec{q})$ the Fourier transform of the bare interaction potential appearing in \eqref{eq:sm_genforce}. 
Given the relationship between $G_L$ and the response function $\chi(\vec{q},\omega)$ expressed in Fig.~\ref{fig:1_SM}, the set of equations is then closed after relating $S(\vec{q})$ to the imaginary part of the response function via the fluctuation-dissipation theorem:
\begin{equation}
   S(\vec{q})=-\frac{\hbar}{\pi n}\int \chi''(q,\omega) d(\omega),
   \label{eq:sm_FlucDiss}
\end{equation}
with $\chi''(q,\omega)$ the imaginary part of the response function. 
If we had to reason within the self-consistent integral-equations methods for the Green's functions, the choice of $G_{L}$ would amount to define the irreducible interaction which, along with the single-particle Green's function, determines the vertex-function correction. This in turn, closing the self-consistent loop, leads the proper response, the effective potential and the single-particle self-energy and Green's function.  
\section{Calculation of the Interaction Correction to the Grand-Canonical Potential}
The interaction correction $\delta\Omega$ to the non-interacting grand-canonical potential is calculated via the ring diagrams in fig.2 in the main text. The resulting expression reads:
\begin{equation}
        \delta\Omega=T\sum_{\mathbf{Q}_n}e^{i\nu_n0^+}
        \ln\{ 1-\left[1-\mathcal{G}(\mathbf{Q}_n)\right] U_{\text{eff}}(\mathbf{Q}_n)\Pi_0(\mathbf{Q}_n)\},
    \label{eq: thermo pot}
\end{equation}
where $\mathbf{Q}_n\equiv (\mathbf{q},i\nu_n)$, $\nu_n=2\pi n T$ is a bosonic Matsubara frequency. The $e^{i\nu_n0^+}$ factor is needed for the Matsubara sum convergence, and as usual $\mathcal{G}$, $\Pi_0$ and $D_0$ are calculated on the imaginary axis. The equation for the density $n$ results
\begin{equation}
    \begin{split}        n=\sum_{\mathbf{k},\sigma}f(\varepsilon_{\mathbf{k}})+2\sum_{\mathbf{q}}n_B(\varepsilon^B_{\mathbf{q}})-\frac{\partial(\delta\Omega)}{\partial \mu},
    \end{split}
    \label{eq: mu}
\end{equation}
where $n_B$ labels the Bose distribution and the factor 2 accounts for each boson being composed of two fermions. 
\section{Renormalization of the Couplings}
In order to make the non-interacting pairing suscptibility $\Pi_0$ convergent, one has to renormalize the couplings. The simplest choice reads \cite{Kokkelmans_Renorm,OhashiGriffinTrap}:
\begin{equation}
U_{bg,R}=\frac{U_{bg}}{1-U_{bg}\gamma};\,\ 
g_R=\frac{g}{1-U_{bg}\gamma};\,\ 
2\nu_R=2\nu-\frac{g^2\gamma}{1-U_{bg}\gamma},
\end{equation}
with $\gamma\equiv \sum_{\mathbf{k}}{m}/{k^2}$.
This simple choice is sufficient to cancel the divergence of the integrals defining $\Pi_0(\mathbf{q},\omega)$, $\Pi_0(\mathbf{q},\mathbf{q}';\omega)$. In the following, we drop $U_{bg}$ for simplicity, though it can easily be restored.

\section{Equations in the Superfluid state with finite Background Scattering Length}
In order to focus on the essence of the theory, in the main text we have derived the equations for the critical temperature $T_c$, where the complex formalism is slightly reduced. Here, we extend the theory to the superfluid state. The corresponding theory equations can be obtained in a similar manner as those for $T_c$: the main difference consists in adding to Hamiltonian (1) in the main text a "mean-field" term containing the pairing gap.\\
We start by performing a mean-field decomposition and Nambu transformation, i.e. 
\begin{equation}
\begin{split}
&\Psi_{\bm{k},\uparrow}=c_{\bm{k},\uparrow}\qquad\qquad\qquad
\Psi_{\bm{k},\downarrow}=c^{\dagger}_{-\bm{k},\downarrow}\\
&\Psi^{\dagger}_{\bm{k},\uparrow}=c^{\dagger}_{\bm{k},\uparrow}\qquad\qquad\qquad
\Psi^{\dagger}_{\bm{k},\downarrow}=c_{-\bm{k},\downarrow},
\end{split}
\label{eqS: Nambu transf}
\end{equation}
on the Boson-Fermion Hamiltonian (1) in the main text. We get 
\begin{equation}
\begin{split}
\mathcal{H}_{\text{BF}}=\frac{\widetilde{\Delta}^2}{U_{\text{eff}}}+\sum_{\bm{k}} \varepsilon_{\bm{k}}+\sum_{\bm{k}}\Psi^{\dagger}_{\bm{k},\sigma}\, h_{\bm{k}}^{\sigma\sigma'}\,\Psi_{\bm{k},\sigma'}+
\sum_{\bm{q}}\varepsilon_{\bm{q}}^{B}\,b^{\dagger}_{\bm{q}}\,b_{\bm{q}}+\frac{g}{\sqrt{V}}\sum_{\substack{\bm{q}}}\left(b^{\dagger}_{\bm{q}}\,\,\hat{\rho}^{-}_{\bm{-q}}+b_{\bm{q}}\,\,\hat{\rho}^{+}_{\bm{q}}\right)
-\frac{U_{bg}}{V}\sum_{\bm{q}}\hat{\rho}^{+}_{\bm{q}}\hat{\rho}^{-}_{\bm{-q}}.
\label{eqS:bf4respfunc}
\end{split}
\end{equation}
Here, we have made use of the definitions
\begin{equation}
\hat{\varrho}_{\bm{q}}^\pm=\sum_{\bm{k},\sigma,\sigma'}\Psi^{\dagger}_{\bm{k}+\bm{q}/2,\sigma}\,\tau^\pm_{\sigma,\sigma'}\,\Psi_{\bm{k}-\bm{q}/2,\sigma'}\quad\text{with}\quad \tau^\pm=\frac{1}{2}\left(\tau^1\pm i \tau^2\right),
\label{eqS: gen. densities}
\end{equation}
with $\tau^i$ the Pauli matrices, $h_{\bm{k}}=\varepsilon_{\bm{k}}\tau^3-\widetilde{\Delta}\tau^1$ ($\widetilde{\Delta}$ is the pairing gap) and $U_{\text{eff}}=U+g^2/(2\nu-2\mu)$. We then define the generalized Wigner distribution functions (in Nambu basis):
\begin{equation}
    f^{\sigma \sigma'}_{\bm{p},\bm{q}}(t)=\langle \Psi^\dagger_{\bm{p}+\bm{q}/2,\sigma}(t)\Psi_{\bm{p}-\bm{q}/2,\sigma'}(t)\rangle,
\end{equation}
and add to the Hamiltonian \eqref{eqS:bf4respfunc} a source term of the form
\begin{equation}
    \sum_{\bm{q},\sigma,\sigma'} J^{\sigma'\sigma}_{-\bm{q}}(t)\sum_{\bm{k}}\Psi^\dagger_{\bm{k}+\bm{q}/2,\sigma}\Psi_{\bm{k}-\bm{q}/2,\sigma'}.
\end{equation}
Now, we then write down the equations of motion for the Wigner distribution functions by commuting them with the perturbed Hamiltonian. We linearize them at first order in the sources, Fourier transform in time, and eventually obtain the expression
\begin{equation}
\begin{split}
\omega\,\delta f^{\sigma\sigma'}_{\bm{p},\bm{q}}&=h^{\sigma s}_{\bm{p}+\bm{q}/2}\,\delta f^{s\sigma'}_{\bm{p},\bm{q}}-\delta f^{\sigma s}_{\bm{p},\bm{q}}\,h^{s \sigma'}_{\bm{p}-\bm{q}/2}+\left(J_{\bm{q}}^{\sigma s}(\omega)\,n^{s\sigma'}_{\bm{p}-\bm{q}/2}-
n^{\sigma s}_{\bm{p}+\bm{q}/2}J_{\bm{q}}^{s\sigma' }(\omega)\right)+\\
&+\sum_{\bm{q}',\eta=\pm}U_{\text{eff}}(-\bm{q}',\omega)\Big\langle\hat{\varrho}_{-\bm{q}'}^{\eta}\,\tau^{-\eta}_{s\sigma }\,\hat{f}^{s\sigma'}_{\bm{p}+\bm{q}'/2,\bm{q}+\bm{q}'}-
\hat{\varrho}_{-\bm{q}'}^{\eta}\,\tau^{-\eta}_{\sigma's }\,\hat{f}^{\sigma s}_{\bm{p}-\bm{q}'/2,\bm{q}+\bm{q}'}\Big\rangle.
\end{split}
\label{eqS:EOM linearized}
\end{equation} 
Here, a summation over repeated indexes is intended. In addition,  $n_{\bm{k}}^{\sigma\sigma'}\delta_{\bm{q},0}=\langle\Psi^\dagger_{\bm{k}+\bm{q}/2,\sigma}\Psi_{\bm{k}-\bm{q}/2,\sigma'}\rangle_0$, with the average performed on the equilibrium system, and $U_{\text{eff}}(\bm{q},\omega)=-U_{bg}+g^2D_0(\bm{q},\omega)$, with $D_0(\bm{q},\omega)=[\omega-\varepsilon_{\bm{q}}^B]^{-1}$ the non-interacting boson propagator.
The last term in eq. \eqref{eqS:EOM linearized} can be decomposed into a connected and an unconnected average, i.e. $\langle(1-c_{\uparrow}^{\dagger}c_{\uparrow}-c_{\downarrow}^{\dagger} c_{\downarrow})\, \hat{\varrho}^+\rangle=\langle(1-c_{\uparrow}^{\dagger}c_{\uparrow}-c_{\downarrow}^{\dagger} c_{\downarrow})\,\hat{\varrho}^+\rangle_C+\langle1-c_{\uparrow}^{\dagger}c_{\uparrow}-c_{\downarrow}^{\dagger} c_{\downarrow}\rangle\langle \hat{\varrho}^+\rangle$. Neglecting the first (connected average) term would lead to the RPA form of the pairing susceptibility, valid for small $g$ and $U$ values. \\ 
If we were in the normal state described in the main text to calculate the critical temperature $T_c$, one would get physical insight by transforming eq. (3) in the main text into the equations of motion for the $f^{(*)}(\bm{R},\bm{r})=\langle c_{\downarrow(\uparrow)}^{(\dagger)}(\bm{R}+\bm{r}/2)c_{\uparrow(\downarrow)}^{(\dagger)}(\bm{R}-\bm{r}/2) \rangle$ in real space. 
Then, one would plug in the STLS approximation to the connected average, i.e.  
\begin{equation}
\sum_{\alpha=\pm}\langle\delta^3(\bm{r})/2-        c_{\sigma(\alpha)}^{\dagger}(\bm{x}_{\alpha})
        c_{\sigma(\alpha)}(\bm{x}_{-\alpha})\rangle\ 
        g_{\text{corr}}\left({r}/{2}\right)\langle \hat{\varrho}^+(\bm{x}_{-\alpha})\rangle,   \label{eq: STLS approx} 
\end{equation}
with $\bm{x}_{\alpha}\equiv  \bm{R}+{\alpha\bm{r}}/{2}$, and transform back to momentum space, getting the linearized and decoupled form of the equations of motion (3) in the main text:
\begin{equation}
    \begin{split}
        &\omega f^*_{\bm{k},\bm{q}}=\left(\varepsilon_{\bm{k}+\bm{q}/2}+\varepsilon_{\bm{k}-\bm{q}/2}\right)f^*_{\bm{k},\bm{q}}\\
        &+(J^*_{\bm{q}}+U_{\text{eff}}(\bm{q},\omega)\langle\hat{\varrho}^+_{\bm{q}}\rangle)\left(1-n_{\bm{k}+\bm{q}/2}-n_{\bm{k}-\bm{q}/2}\right)\\
    &+\langle\hat{\varrho}^+_{\bm{q}}\rangle\sum_{\bm{q}'}U_{\text{eff}}(\bm{q}',\omega)S(\bm{q}-\bm{q}')\left(1-n_{\bm{k}+\bm{q}'/2}-n_{\bm{k}-\bm{q}'/2}\right).
    \end{split}
    \label{eq: EOM wigner final}
\end{equation}
The presence of the static $S(\bm{q})$ arises from the equilibrium value of $\langle c\,c\,\varrho^+\rangle$ in the linearized equation of motion and clarifies the choice of $g_{\text{corr}}$ for decomposing the connected average as in the STLS approximation. Neglecting the interaction terms in~\eqref{eq: EOM wigner final} leads to the asymptotic limits for the local field factor, exactly as performed by Niklasson \cite{Niklasson} and Zhu and Overhauser \cite{ZhuOverh} for the electron liquid.\\

Going back to the general case~\eqref{eqS:EOM linearized}, we write the connected average in real space basis and we approximate it with the help of \emph{generalized pair correlation functions}:
\begin{equation}
\begin{split}
\Big\langle\hat{\varrho}^{\eta}(\bm{x}'')\,\tau^{-\eta}_{s\sigma }\,\hat{\varrho}^{s\sigma' }(\bm{x},\bm{x}')\Big\rangle_C\simeq& 
\sum_{\alpha=\pm}\varrho^{\eta}(\bm{x}'')\,g_{\text{corr}}^{\alpha;-\alpha}(|\bm{R}-\bm{x}''|)\,\tau^{-\eta}_{s\sigma}\,\varrho^{s\sigma'}(\bm{x},\bm{x}').
\label{eqS: STLS approximation}
\end{split}
\end{equation}
Here,  
\begin{equation}
g_{\text{corr}}^{\eta;\eta'}\left(\left|\bm{x}-\bm{x}''\right|\right)\equiv
\Big\langle\hat{\varrho}^{\eta}(\bm{x}'')\,\hat{\varrho}^{\eta'}(\bm{x},\bm{x})\Big\rangle_{0,C},
\label{eq4:gF}
\end{equation}
where the subscript $0,C$ means that the connected average is performed over the unperturbed system. We then transform back the equations of motion in momentum space and decompose the Wigner distribution function in its Pauli components:
\begin{equation}
    f^i_{\bm{p},\bm{q}}=\sum_{\sigma\sigma'}\tau^i_{\sigma'\sigma}f^{\sigma\sigma'}_{\bm{p},\bm{q}}.
\end{equation}
Each Pauli component has a precise physical meaning: indeed, components 1 and 2  describe amplitude and phase fluctuations of the order parameter, respectively, while components 3 and 4 are associated with density and spin fluctuations. After performing the same transformation on the sources $J$, we notice that the response of the system to the different sources $J^i$ will be given by a $4\times 4$ matrix response function $\Pi$. Inverting the equations of motion, we then get an expression for $\Pi$ (all multiplications and divisions must be intended in a matrix sense): 
\begin{equation}
\Pi(\bm{q},\omega)=\frac{\Pi_0(\bm{q},\omega)}{1-\Pi_0(\bm{q},\omega)\left\{\Hat{U}_{R,\text{eff}}(\bm{q},\omega)\left[W-\mathcal{G}(\bm{q},\omega)\right]\right\}},
\label{eqS: STLS response function}
\end{equation}
with $W=\text{diag}\{1,1,0,0\}$ and $\Hat{U}_{R,\text{eff}}(\bm{q},\omega)$ the (rotated) matrix effective interaction $\text{diag}\{-U+g^2D_0(\bm{q},\omega),-U+g^2D_0(-\bm{q},-\omega),0,0\}$. In particular, this is rotated by the unitary matrix
\begin{equation}
    R=\left(
    \begin{array}{cccc}
         1/\sqrt{2}&1/\sqrt{2}&0&0   \\
         -i/\sqrt{2}&i/\sqrt{2}&0&0   \\
         0&0&1&0\\
         0&0&0&1
    \end{array}
    \right).
\end{equation}

$\Pi_0(\bm{q},\omega)$ is the response function calculated only with the mean-field Hamiltonian, whose expressions can be found, for example, in~\cite{Ohashi_Griffin_goldstone}. It turns out that all the components of $\Pi_0$ connecting a spin fluctuation with another kind of excitation are zero. Indeed, by means of symmetry principles, it can be proved at all orders that spin fluctuations completely decouple from the others \cite{effective_action-Castellani}. 
The local field factor is now a matrix defined as 
\begin{equation}
\mathcal{G}(\bm{q},\omega)\equiv-\sum_{\bm{q}'}\frac{W\,\Pi_{0}(\bm{q},\bm{q}',\omega;\widetilde{\Delta}=0)\,W}{W\,\Pi_{0}(\bm{q},\omega;\widetilde{\Delta}=0)\,W}\,\left[S^{11}(\bm{q}-\bm{q}')+S^{22}(\bm{q}-\bm{q}')\right],
\label{eqS: LocalFieldFactor}
\end{equation}
where $S^{11}$ and $S^{22}$ are the amplitude-amplitude and phase phase structure factors, respectively. They are related to the 11 and 22 components of the response functions by the fluctuation-dissipation theorem. Inside the summation in eq. \eqref{eqS: LocalFieldFactor}, the quantity $\Pi_{0}(\bm{q},\bm{q}',\omega;\widetilde{\Delta}=0)$ appears. The dependence of the latter on two momenta has the following meaning: since $\Pi_0$ can be expressed as the integral of a fraction, we assign two different momenta in the numerator and denominator, similarly to what we have done in the main text. Finally, since approximation \eqref{eqS: STLS approximation} is expected to be valid in the high frequency limit, we can neglect $\widetilde{\Delta}$ ($\omega\gg\widetilde{\Delta}$) in the calculation, so that $W\,\Pi_{0}(\bm{q},\bm{q}',\omega;\widetilde{\Delta}=0)\,W$ and $W\,\Pi_{0}(\bm{q},\omega;\widetilde{\Delta}=0)\,W$ precisely equal to their analog in the normal state. We have constrained this approximation to provide a response function satisfying all the fundamental spin-flip and time-reversal symmetries of the Hamiltonian \eqref{eqS:bf4respfunc}~\cite{PMB_Thesis}.\\ 
The renormalized boson propagator is:
\begin{equation}
D(\bm{q},\omega)=\frac{D_{0}(\bm{q},\omega)}{1-D_{0}(\bm{q},\omega)\,\Sigma_{\text{STLS}}(\bm{q},\omega)}, 
\end{equation}
where the self-energy equals
\begin{equation}
\begin{split}
\Sigma_{R,\text{STLS}}(\bm{q},\omega)=g^2\,R^{-1}\left[W\frac{\Pi_0(\bm{q},\omega)\left[W-\mathcal{G}(\bm{q},\omega)\right]}{1+U\Pi_0(\bm{q},\omega)\left[W-\mathcal{G}(\bm{q},\omega)\right]}W\right]R
.\end{split}
\label{eqS: STLS self energy}
\end{equation}
By requiring that the Hugenholtz and Pines theorem be fulfilled~\cite{Hugenh-Pines}, i.e. by requiring that the bosonic spectrum be gapless and 
\begin{equation}
    \det D^{-1}(\bm{q}=0,\omega=0)=0,
\end{equation}
we get the renormalized equation for the superfluid gap:
\begin{equation}
1=U_{\text{eff}}^{\text{STLS}}(0,0)\sum_{\bm{p}}\frac{\tanh\left(\frac{1}{2}\beta E_{\bm{p}}\right)}{2E_{\bm{p}}}.
\label{eqS: gap equation STLS}
\end{equation}
with $U_{\text{eff}}^{\text{STLS}}(0,0)=[U+g^2/(2\nu-2\mu)][1-\mathcal{G}^{11}(0,0)]$. 
The number equation is then obtained as in the main text, by summing up all closed ring diagrams:
\begin{equation}
N=2\phi_m^2+\sum_{\bm{p}}\left[1+\frac{\varepsilon_{\bm{p}}}{E_{\bm{p}}}\tanh\left(\frac{\beta}{2}E_{\bm{p}}\right)\right]+2\sum_{\bm{q}}n_B(\varepsilon_{\bm{q}})
+\frac{1}{\beta}\sum_{\bm{q},i\nu_n}\frac{\partial}{\partial\mu}{\rm tr}\ln\left[1+\Pi_0(\bm{q},i\nu_n)\,\Hat{U}_{\text{R,eff}}^{\text{STLS}}(\bm{q},i\nu_n)\right]e^{i\nu_n0^+},
\label{eqS: number equation STLS}
\end{equation}
with $\phi_m^2$ being the fraction of condensed molecules, $E_{\bm{p}}=\sqrt{\varepsilon_{\bm{p}}^2+\widetilde{\Delta}^2}$ the quasiparticle dispersion, and
\begin{equation}
    \Hat{U}_{\text{R,eff}}^{\text{STLS}}(\bm{q},\omega)=\left[W-\mathcal{G}(\bm{q},\omega)\right]\Hat{U}_{\text{R,eff}}(\bm{q},\omega).
\end{equation}

\bibstyle{apsrev}